\documentclass{jg_symp}
\usepackage{psfig}

\def\ASCA{{\it ASCA}}
\def\ROSAT{{\it ROSAT}}
\def\Einstein{{\it Einstein}}
\def\SAX{{\it SAX}}
\def\etal{et al.}
\def\NH{$N_\mathrm{H}$}
\def\NHUNIT{H~cm$^{-2}$}
\def\FLUXUNIT{erg~s$^{-1}$~cm$^{-2}$}
\def\LUMIUNIT{erg~s$^{-1}$}
\begin{document}
\title{SNRs in the Galactic Center region observed with {\it ASCA}}

\author{M. Sakano\inst{1,3}, J. Yokogawa\inst{1}, H. Murakami\inst{1}, K. Koyama\inst{1,5} Y. Maeda\inst{2,4} \and The {\ASCA} Galactic Plane/Center Survey team}
\institute{Department of Physics, Kyoto University, Sakyo, Kyoto 606-8502, Japan
\and  Department of Astronomy and Astrophysics,
 Pennsylvania State University,
 University Park PA 16802-6305 U.S.A.
\and  Research Fellow of the Japan Society for the Promotion of Science
\and  Postdoctctoral Fellow for Research Abroad of JSPS
\and  CREST: Japan Science and Technology Corporation (JST)}

\authorrunning{Sakano et al.}
\titlerunning{SNRs in the Galactic Center region observed with {\it ASCA}}

\maketitle

\begin{abstract}
    We report the {\ASCA} results of the supernova remnants (SNRs)
 and their candidates in the Galactic Center region.
    We found apparent X-ray emission from G359.1$-$0.5 and G0.9$+$0.1,
 and made marginal detection for G359.1$+$0.9,
 but found no significant X-ray from the other cataloged SNRs: G359.0$-$0.9, 
 Sgr A East (G0.0$+$0.0), G0.3$+$0.0, Sgr D SNR (G1.0$-$0.1) (\cite{Green98}).
    The emission from G359.1$-$0.5 is found to be thermal with
 multi temperature structures
 whereas that from G0.9$+$0.1 is quite hard, probably non-thermal.
    We discovered two new candidates of SNRs: G0.0$-$1.3 (AX~J1751$-$29.6)
 and G0.56$-$0.01 (AX~J1747.0$-$2828).  The former, G0.0$-$1.3, shows
 the extended emission with a thin thermal plasma.  The latter,
 G0.56$-$0.01, shows quite strong 6.7-keV line with the equivalent width
 of 2 keV, which resembles that of the Galactic Center plasma.
    We discuss the nature of those SNRs, relating with the origin
 of the Galactic Center hot plasma.

\end{abstract}

\section{Introduction}

    The hard X-ray structure of the Galactic Center region is quite unusual
 as well as those of the other wavelengths.
  In particular, thin thermal hot plasma
 prevails all over the region (\cite{Koyama89}; \cite{Yamauchi90};
 \cite{Koyama96}; \cite{Maeda98}).
  It has several surprising natures.  First, the temperature is quite high,
 about 10~keV.
  Second, the total energy is also high, about $10^{54}$ erg.
  Third, the plasma wide-spreads in the region bigger than 1\degr$\times$1\degr,
 which corresponds to about 150pc$\times$150pc for the distance of
 the Galactic Center.
  Fourth, the spectrum of the plasma is surprisingly uniform from field
 to field except for the slight change of the surface brightness.
  The origin of the plasma is still an enigma.

    Supernova remnants (SNRs) are one of the possible candidates for the origin
 of the hot plasma (e.g., \cite{Koyama86}), as well as the other candidates;
 the past activity of the central massive black hole
 (e.g., \cite{Koyama96}), unresolved many cataclysmic
 variables (e.g., \cite{Mukai93}), and 
 the global magnetic activity which heats the interstellar matter
 (e.g., \cite{Yokoyama98}).

    If the SNRs explain all the hot plasma, first, the total energy
 ($\sim 10^{54}$ erg) requires about 10$^3$ supernovae in the narrow region
 with the scale of a few hundred pc in the last 5\,10$^4$ years,
 which is the age of the plasma (\cite{Koyama96}).
  It would lead to the much higher supernova rate than the common rate
 of a few supernovae in a century in all the Galaxy.  However,
 the possible starburst activity in the Galactic Center region has been
 often discussed.
  In particular, recent {\it COMPTEL} observations
 of \element[][26]{Al} 1.8~MeV line suggest that 10$^5$ supernovae occurred
 in the last 4\,10$^6$ years (\cite{Chen95}; \cite{Hartmann95}),
 which is consistent with the required supernova rate.

    Second, the supernova origin hypothesis
 requires the mean temperature of about 10 keV.
  It is much higher than the usual temperature of SNRs
 ($kT\leq$ a few keV).   However, if SNRs densely
 exist in the region, the temperature may heat up to about 10 keV
 by the mutual interaction of the shocks.
    Thus, SNRs are still one of the candidates for the origin
 of the Galactic Center plasma.

    In this paper we concentrate on the X-ray natures of the SNRs
 in the Galactic Center region from the observational point of view.
    The {\ASCA} capability of imaging spectroscopy with 0.5--10 keV (\cite{Tanaka94})
 is quite suitable for the study.   In fact, owing to
 the hard X-ray sensitivity of {\ASCA}, particularly above 2~keV,
 we can possibly detect new SNRs which have been hidden
 by the heavy interstellar absorption, hence obtain some information
 whether the SNRs are valid for the origin of
 the Galactic Center plasma.

    Using all the available {\ASCA} pointing observations in this region
 ($-$1\degr$<l<$1\degr, $-$1\degr$<b<$1\degr)
 and a part of the data of the ongoing {\ASCA} Galactic Center survey,
 we studied the X-ray emission from the known SNRs
 (\cite{Green98}), and searched new SNRs.
    We assume the distance to the Galactic Center to be 8.5 kpc.

\section{Results on the cataloged SNRs}
  
   We searched X-ray emission from the cataloged SNRs (\cite{Green98}),
 and then investigated each nature where significant X-ray emission is detected.

\subsection{G359.1$-$0.5}

\begin{figure}
\centerline{\psfig{file=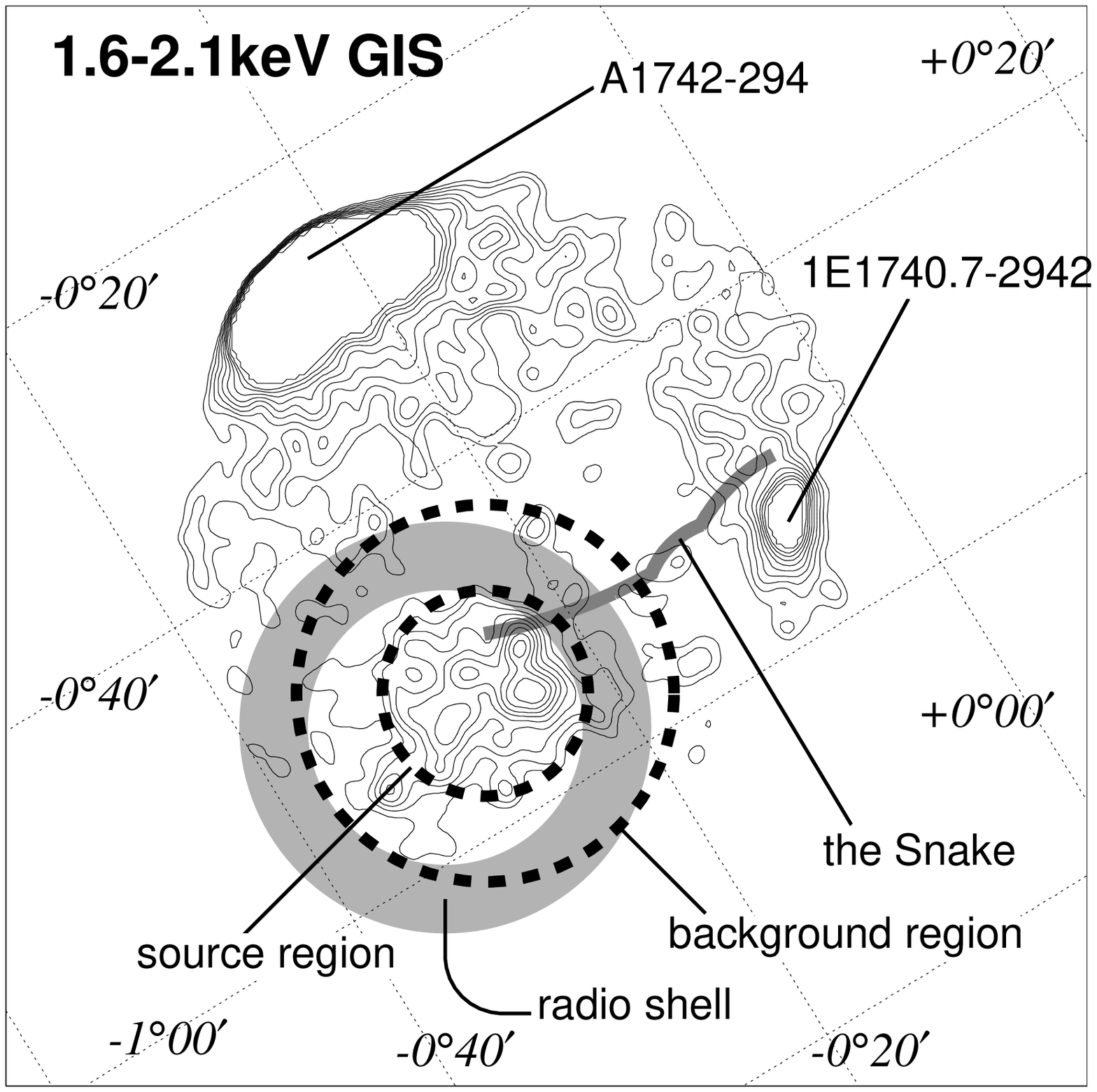,width=8.8cm,clip=} }
\vspace*{0.5cm}
\centerline{\psfig{file=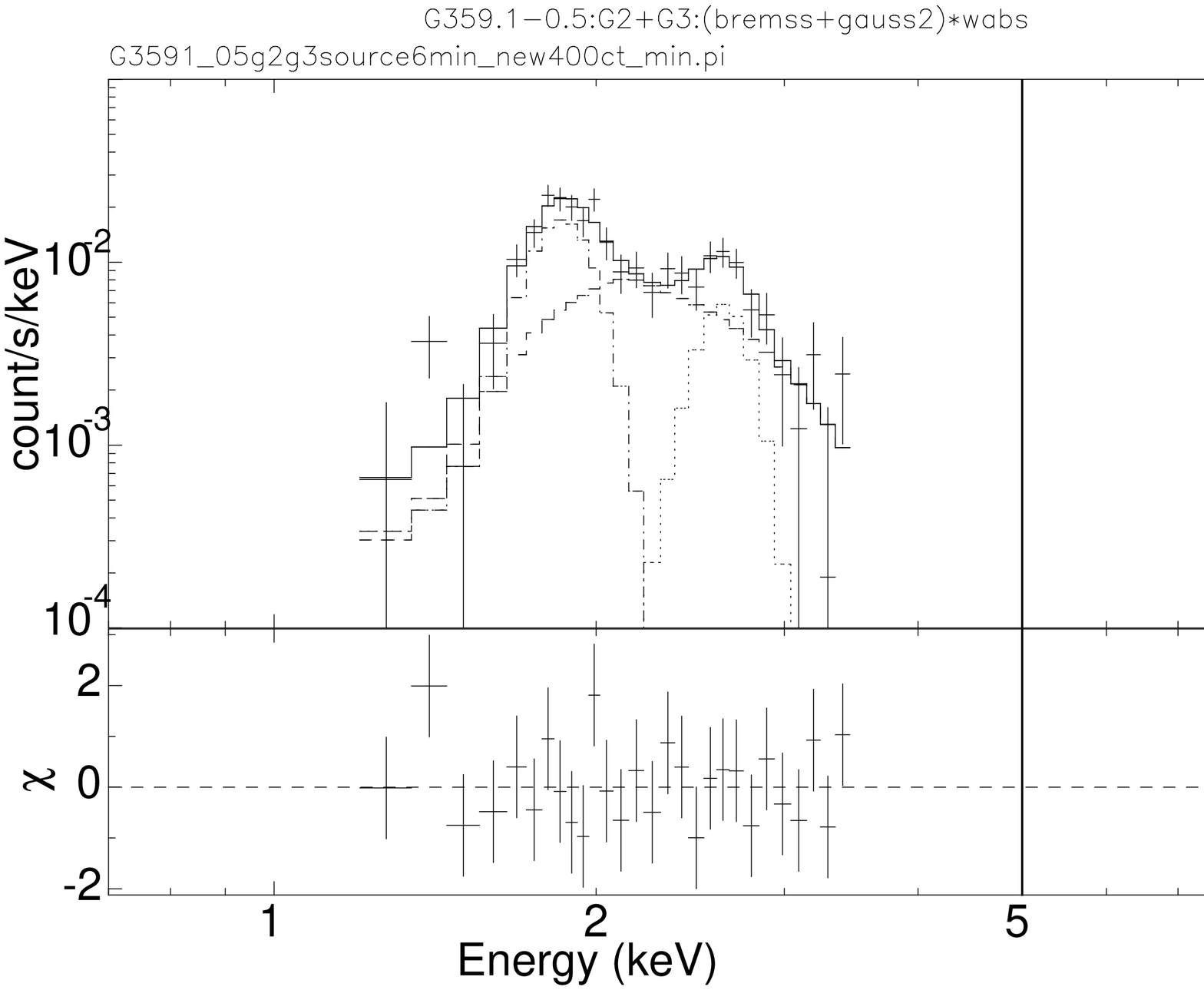,width=8.8cm,clip=} }
\caption{
(Upper) GIS contour map with 1.6--2.1 keV band superposed on a schematic
 diagram of the radio structures;  the radio shell of G359.1$-$0.5
 and the radio non-thermal filament, the Snake.
 Contour level is linearly
 spaced and is saturated for A1742$-$294 and 1E~1740.7$-$2942.
 The accumulated regions for the spectra are also noted.
(Lower) The background-subtracted GIS spectrum.
  We also show the best-fit model where we fit the spectrum with
 the model of the thermal bremsstrahlung and two narrow Gaussians
 with interstellar absorption.
  These two figures are adopted from Yokogawa {\etal} (1999).
\label{fig:359.1-0.5}}
\end{figure}

   Fig.~\ref{fig:359.1-0.5} (upper panel) shows the {\ASCA} GIS image of
 G359.1$-$0.5 with 1.6--2.1 keV band.
   We detected center-filled X-rays from this source, whereas
 the radio image shows a clear shell-like structure (e.g., \cite{Uchida92}).

   The spectrum is found to have emission lines from highly ionized ion
 (Fig.~\ref{fig:359.1-0.5} (lower panel)),
 hence the X-rays come from thin thermal plasma.  The most distinct two lines
 are K$\alpha$ lines from helium-like silicon and hydrogen-like sulfur.
  They imply that the plasma has multi-temperature structure.
  The absorption column density is estimated at \NH$\sim 8\,10^{22}$\NHUNIT,
 suggesting that G359.1$-$0.5 is located at near the Galactic Center,
 which is consistent with the radio observations (\cite{Uchida92}).
  This column density is larger by a factor of 2 or 3 than that with
 the {\ROSAT} measurement (\cite{Egger98}).
  However, we are still confident of our estimation because the {\ASCA}
 energy band is the most suitable for measuring such heavy absorption.
   Details of the analysis are given in Yokogawa {\etal} (1999).

\subsection{G0.9$+$0.1}

\begin{figure}
\centerline{\psfig{file=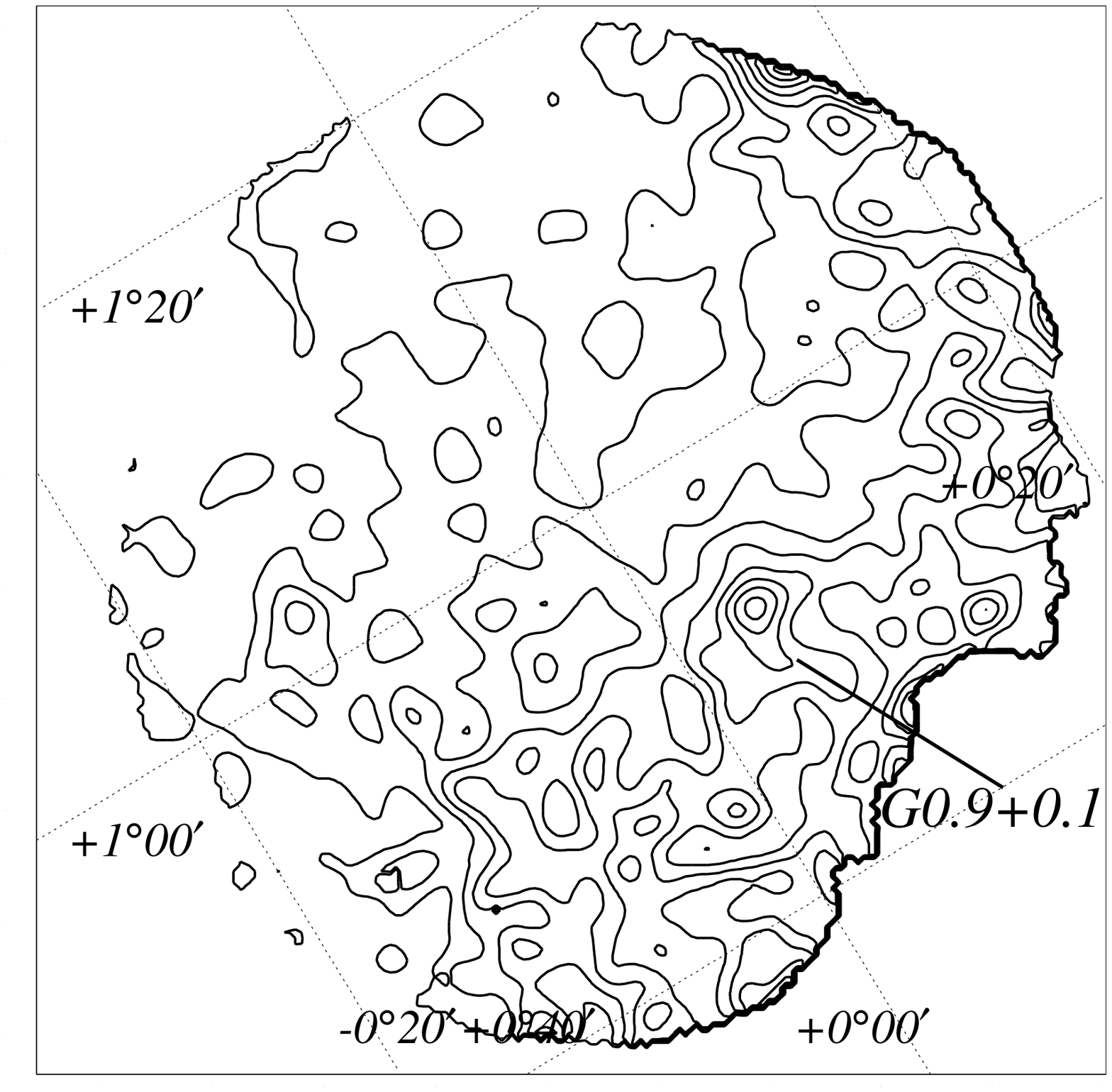,width=8.8cm,clip=} }
\vspace*{-0.5cm}
\centerline{\psfig{file=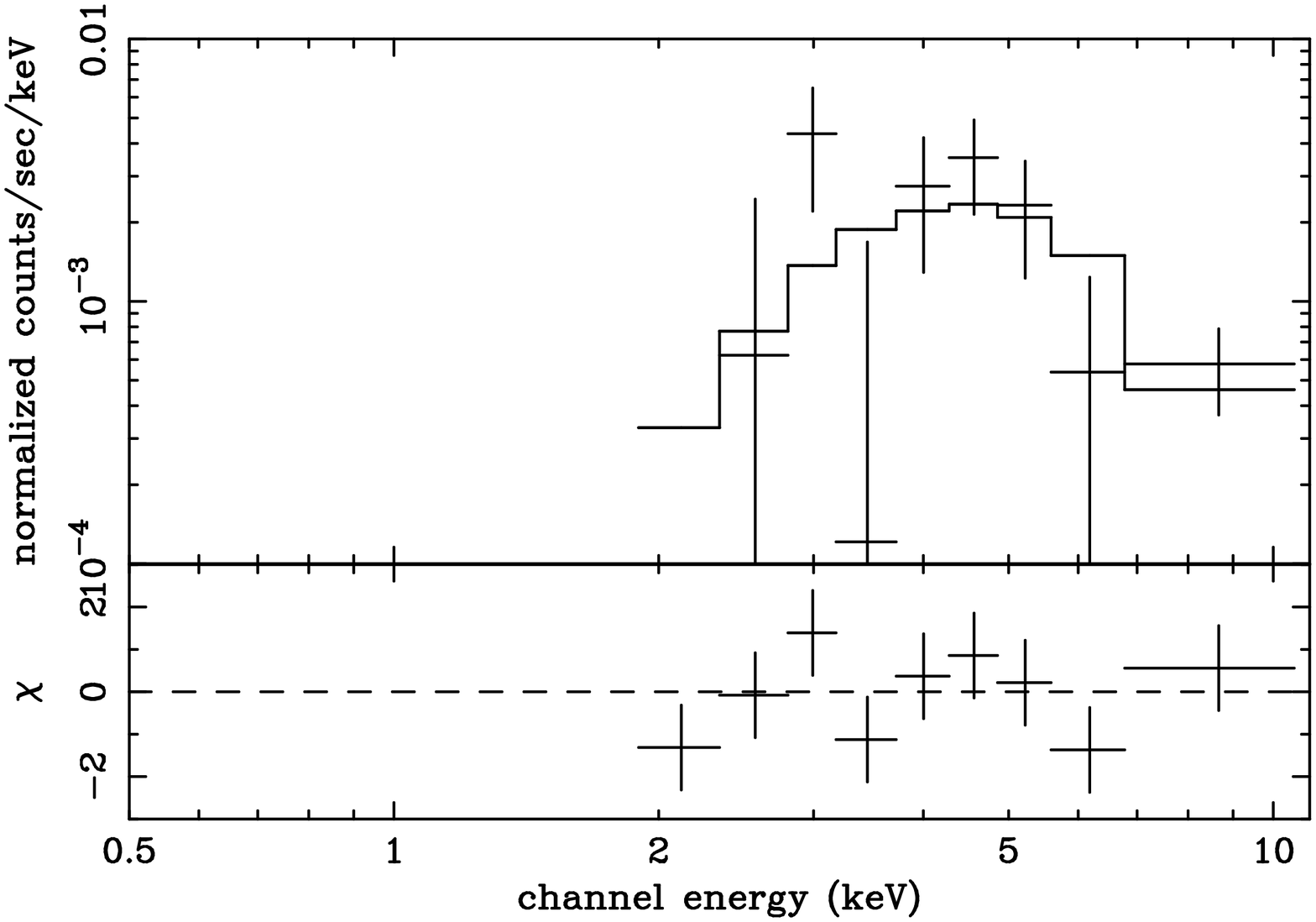,width=8.8cm,clip=} }
\caption{
(Upper) GIS2$+$GIS3 contour map of G0.9$+$0.1 with 3--10 keV band.
 The image is smoothed with a Gaussian filter of $\sigma=0.75$\arcmin,
 and corrected for exposure, vignetting and the GIS grid structure
 after subtraction of non-X-ray background (NXB).  
 Contour level is linearly spaced.
 The coordinate is in galactic ($l_{\mathrm{II}}$, $b_{\mathrm{II}}$),
 and the north is up.
(Lower) The same as Fig.~\ref{fig:359.1-0.5} lower panel, but of G0.9$+$0.1.
  The fitting model is an absorbed power-law.
\label{fig:img:0.9+0.1}}
\end{figure}

   Fig.~\ref{fig:img:0.9+0.1} (upper panel) shows the GIS image of G0.9$+$0.1
 with 3--10 keV band.   We detected significant X-ray emission
 from this source in this hard energy band, whereas no significant X-ray
 is detected in the softer energy band.

   The X-ray emitting region is compact
 and not resolved with {\ASCA} GIS.  The radio size of the source
 is 2{\arcmin} in diameter (\cite{Helfand87}), hence it is consistent
 with the X-ray image.

   The spectrum is found to be hard; $kT>$2 keV (the best-fit of 40 keV)
 in thermal model or $\Gamma\sim$1.5 in power-law model,
 with heavy absorption of \NH$\sim 10^{23}$\NHUNIT,
 although the statistics are not good.  This hardness may imply the emission
 to be non-thermal origin.
   The flux is 2\,10$^{-12}$ {\FLUXUNIT} in 2--10 keV band.
   It is consistent with the upper limit by {\Einstein} (\cite{Helfand87}),
 but significantly lower than the {\SAX} result (\cite{Mereghetti98}).

\subsection{G359.1$+$0.9}

\begin{figure}
\centerline{\psfig{file=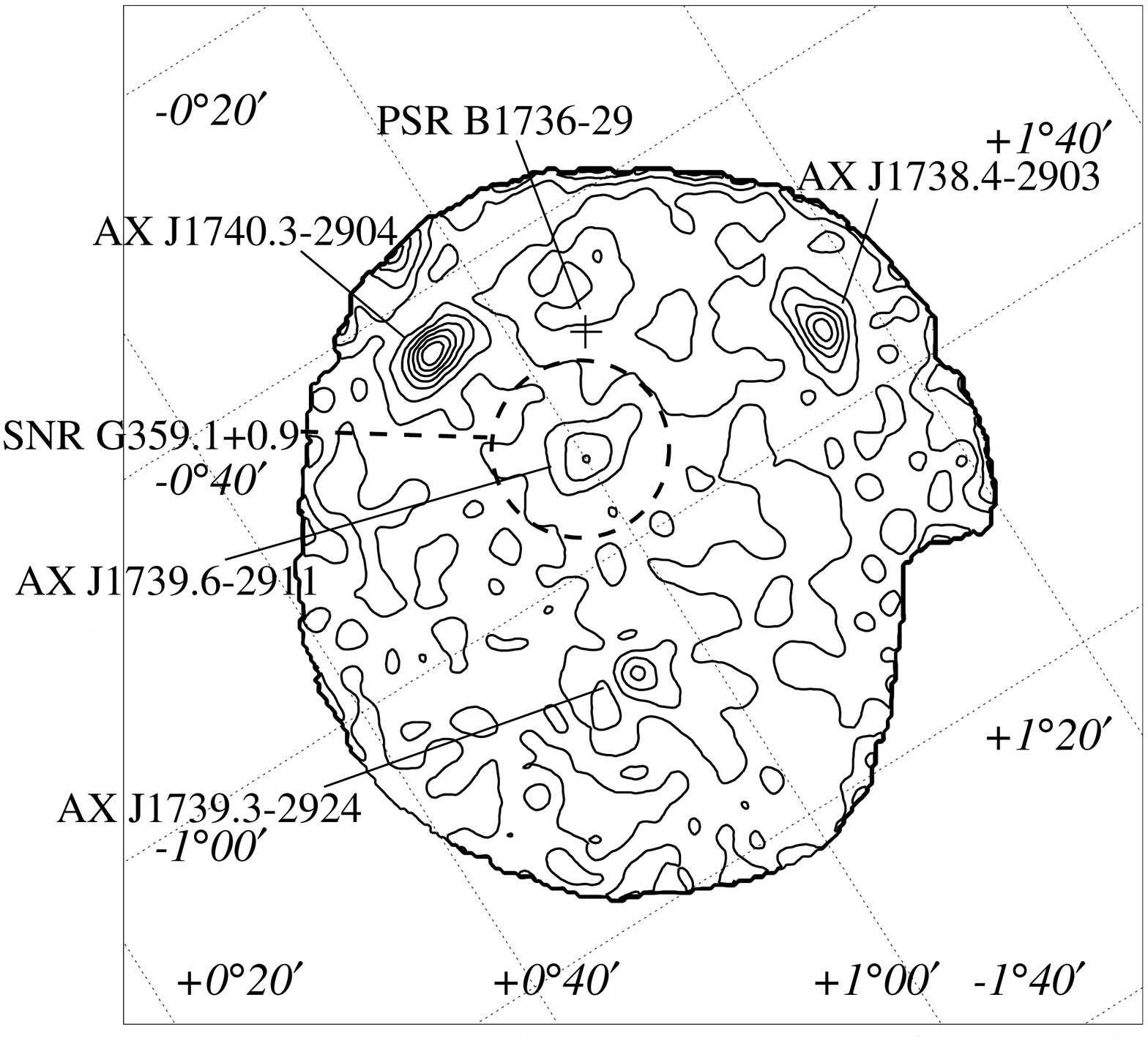,width=8.8cm,clip=} }
\centerline{\psfig{file=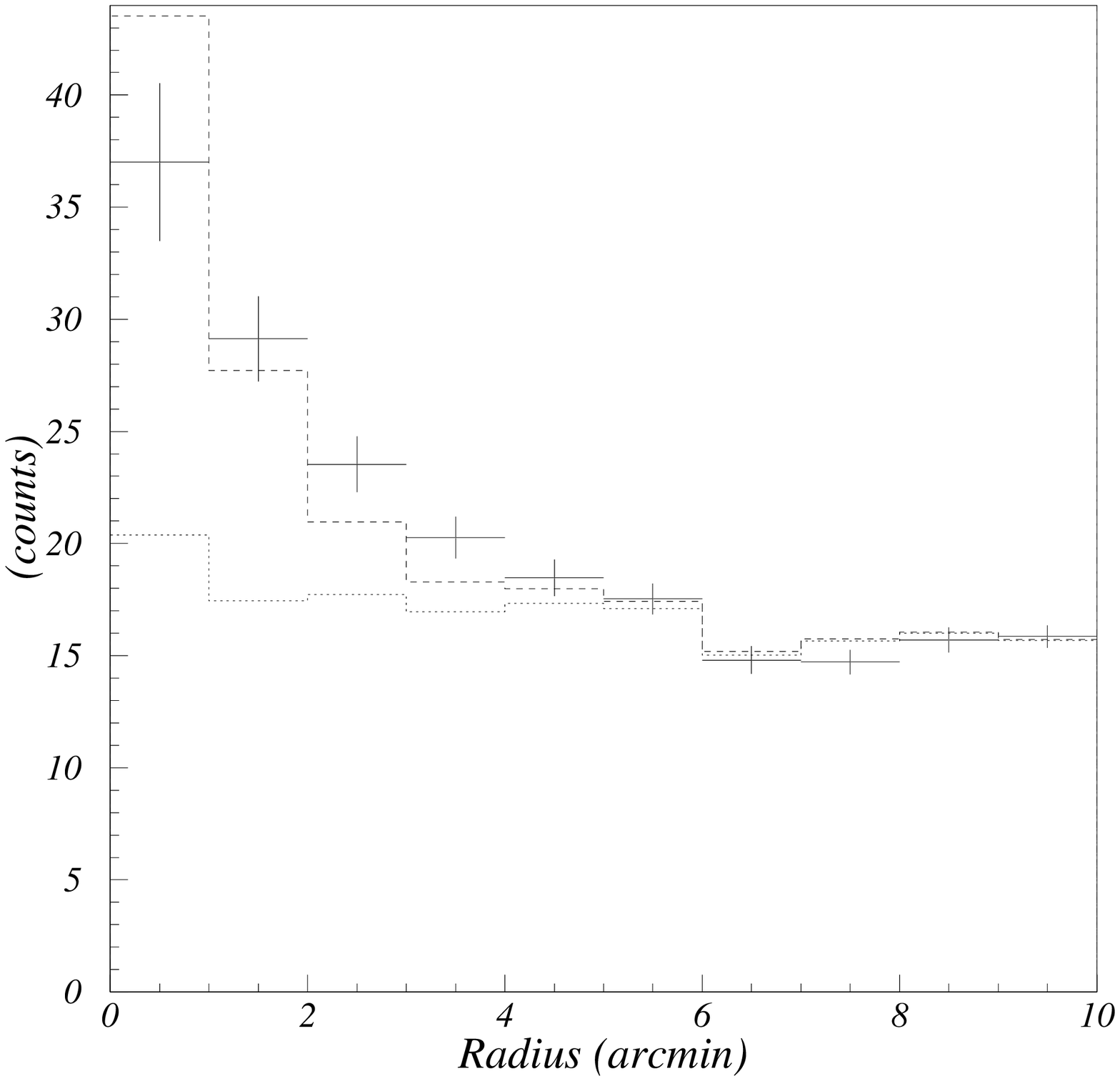,width=8.8cm,clip=} }
\caption{
 (Upper) The same as Fig.~\ref{fig:img:0.9+0.1} upper panel, but of G359.1$+$0.9
 with 0.7--3.0 keV band.
 The positions of the detected X-ray sources
 (AX~J1738.4$-$2903, AX~J1739.3$-$2924, AX~J1739.6$-$2911, AX~J1740.3$-$2904),
 the radio shell of G359.1$+$0.9,
 and the radio pulsar PSR B1736$-$29 are indicated.
 (Lower) The radial profile with the center at the peak of AX~J1739.6$-$2911.
  The profile is fitted with the point spread function (see text).
  The best-fit model is given with the dashed line, whereas the dotted line
 shows only the background component in the best-fit model.
\label{fig:img:359.1+0.9}}
\end{figure}

    Fig.~\ref{fig:img:359.1+0.9} (upper panel) shows the GIS image
 around G359.1$+$0.9.   X-ray emission from the position corresponding
 to G359.1$+$0.9 is detected with the significance of 9.8$\sigma$
 in 0.7--3 keV band (AX~J1739.6$-$2911 in Fig.~\ref{fig:img:359.1+0.9}),
 whereas the significance in 3--10 keV band is 2.1$\sigma$.

    Faint extended emission around AX~J1739.6$-$2911 also can be seen
 in the 0.7--3 keV band image.
  We made the radial profile with the center at the peak of AX~J1739.6$-$2911
 and fitted it with the model of the point spread function (PSF) and
 the background (NXB$+$CXB(cosmic X-ray background)) where the normalizations
 of the PSF and the background were allowed to be free.
  The model is found to be rejected with the confidence of 97.4\%,
 \emph{i.e.}, the slightly extended emission with the radius of
 4{\arcmin}$\sim$5{\arcmin} was marginally detected with the significance of
 2.2$\sigma$ (Fig.~\ref{fig:img:359.1+0.9} lower panel).
  The size is consistent with the radius of the radio shell, $r\sim$5{\arcmin}.

   The spectrum of the central region of AX~J1739.6$-$2911
 is found to be well fitted with the thin thermal plasma model
 with $kT\sim$0.7 keV ($\chi^2$/d.o.f.$=$2.67/5) and \NH$\sim 0$,
 although the statistics are not good.
   The flux is 3\,10$^{-13}$ {\FLUXUNIT} in total energy band,
 converted to the luminosity of 2\,10$^{33}$ {\LUMIUNIT} if we assume
 the distance to be 8.5 kpc.  Note that the soft spectrum with no absorption
 may imply that this source is not located in the Galactic Center region,
 but is a foreground source, hence the above estimated luminosity may
 be reduced by several factors or more.

   The extended emission with positional coincidence with the radio structure
 strongly supports that it is an X-ray counter part of the SNR,
 if the detection of the extended emission is true.
   The soft spectrum also supports it.
   The luminosity may be rather low for an SNR.  However, {\ASCA}
 Plane survey has failed to detect many radio SNRs, suggesting
 their quite dim X-ray luminosities (\cite{Yamauchi99}).  Therefore,
 the low X-ray luminosity of this source may be acceptable for an SNR.
   Future observations with high sensitivity will be encouraged.

\subsection{The other cataloged SNRs}

   For the other cataloged SNRs (G359.0$-$0.9, Sgr A East (G0.0$+$0.0),
 G0.3$+$0.0, Sgr D SNR (G1.0$-$0.1)), we detected no significant X-ray emission
 associated with the SNRs.  It is possibly because no strong X-ray is
 actually emitted.  However we should be conservative for no X-ray emission
 from those sources because the detected positions of those SNRs are
 heavily contaminated from nearby strong X-ray sources;
 SLX~1744$-$299/300 near G359.0$-$0.9,
 Sgr~A and AX~J1745.6$-$2901 near Sgr~A East,
 1E~1743.1$-$2843 near G0.3$+$0.0, and GX3$+$1 near G1.0$-$0.1.

\section{Results on the new candidates of SNRs}

   We discovered two new candidates of SNRs.
  In this section, we report the preliminary results on them.

\subsection{G0.0$-$1.3 (AX~J1751$-$29.6)}

\begin{figure}
\centerline{\psfig{file=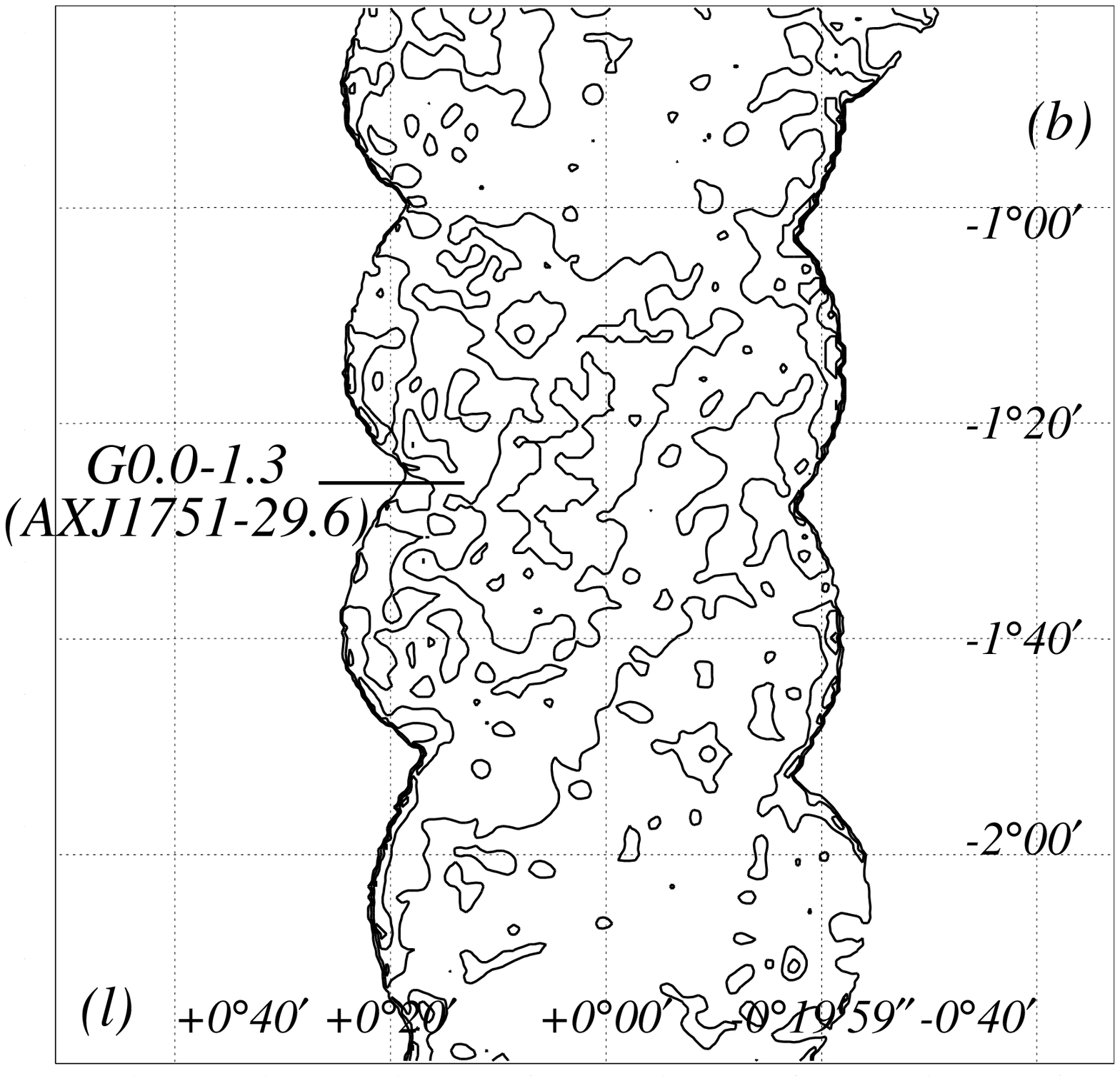,width=8.8cm,clip=} }
\vspace*{-0.5cm}
\centerline{\psfig{file=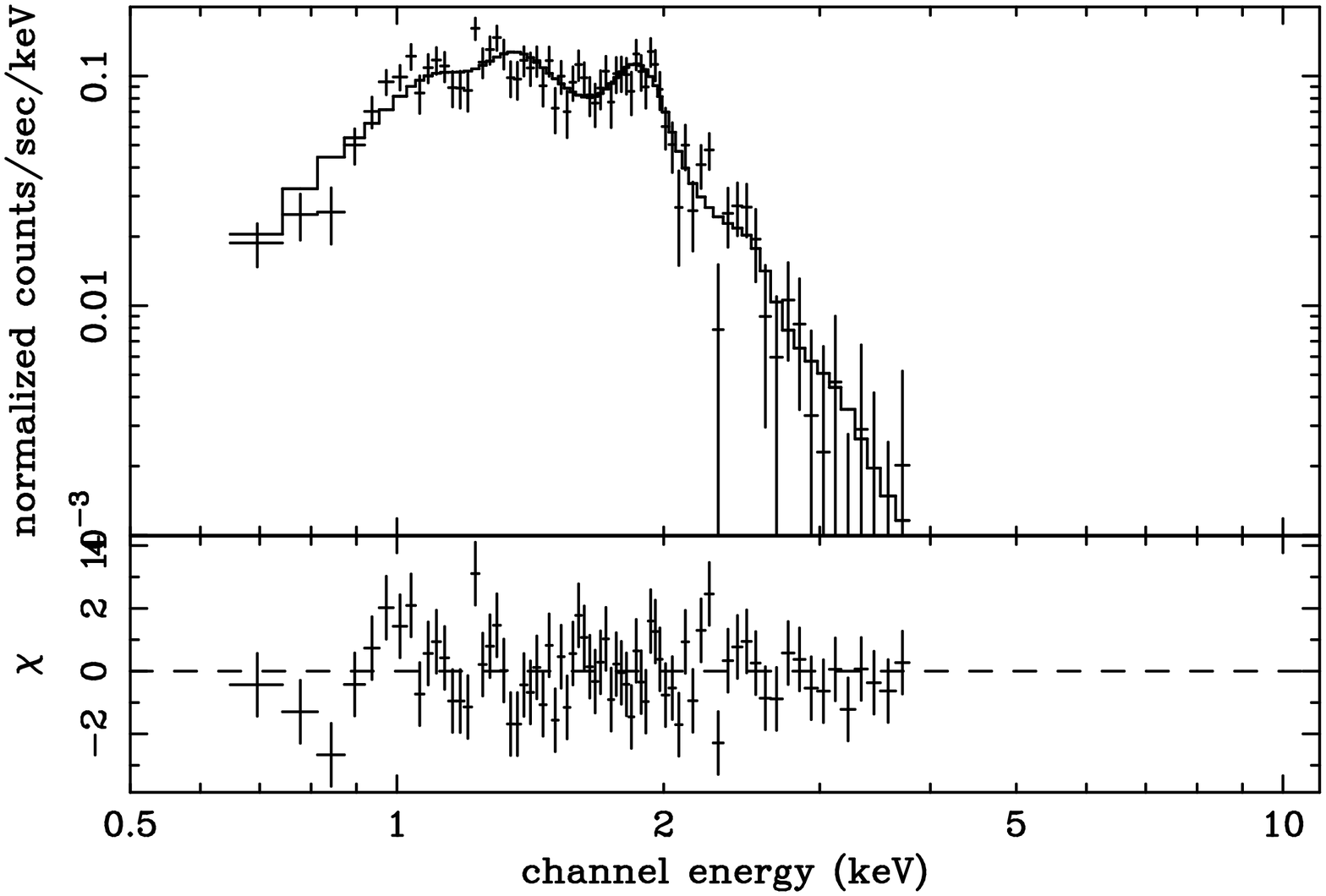,width=8.8cm,clip=} }
\caption{
(Upper) The same as Fig.~\ref{fig:img:0.9+0.1} upper panel, but of G0.0$-$1.3
 (AX~J1751$-$29.6) with 0.7--3.0 keV band.
(Lower) The same as Fig.~\ref{fig:359.1-0.5} lower panel, but of G0.0$-$1.3.
    For the background spectrum, we accumulated the photons from the elliptical
 region surrounding the source, excluding the source region,
 in the same GIS field of view.
   The fitting model is the thin thermal plasma model with absorption.
\label{fig:0.0-1.3}}
\end{figure}

   Fig.~\ref{fig:0.0-1.3} (upper panel) shows the GIS image of G0.0$-$1.3
 (AX~J1751$-$29.6) in 0.7--3 keV band.   We discovered the clearly extended
 emission with the scale of about 40\arcmin$\times$10\arcmin.

   The spectrum is found to have the emission lines from highly ionized ions,
 hence is the thermal origin (Fig.~\ref{fig:0.0-1.3} lower panel).
  In fact, the spectrum is well fitted with a thin thermal plasma model
 with $kT=0.5\pm 0.08$ keV and \NH$= (1.3\pm 0.2)$\,10$^{22}$ {\NHUNIT}.
  The X-ray flux is $\sim 10^{-11}$ {\FLUXUNIT} in 0.5--3 keV band.

  The column density suggests that this source may be in front of
 the Galactic Center region according to Sakano {\etal} (1999),
 and be located at the distance of about 4~kpc if we assume
 the mean interstellar density of 1 H~cm$^{-3}$.
  Then, the obtained flux is converted to the luminosity
 of 3\,$10^{35}$ {\LUMIUNIT} under the assumption of the distance of 4~kpc.
  Even in the case of the quite small distance of 1~kpc, the luminosity
 is larger than $10^{34}$ {\LUMIUNIT}.

  We now consider the classification of the source.
  The clearly extended emission of the thin thermal plasma with $kT\sim$0.5 keV
 implies that this source is an SNR or a star forming region.
  The luminosity is also within the typical range of SNRs but much higher
 than that of a star forming region.
  Therefore, AX~J1751$-$29.6 is a strong candidate for a new SNR.

\subsection{G0.56$-$0.01 (AX~J1747.0$-$2828)}

\begin{figure}
\centerline{\psfig{file=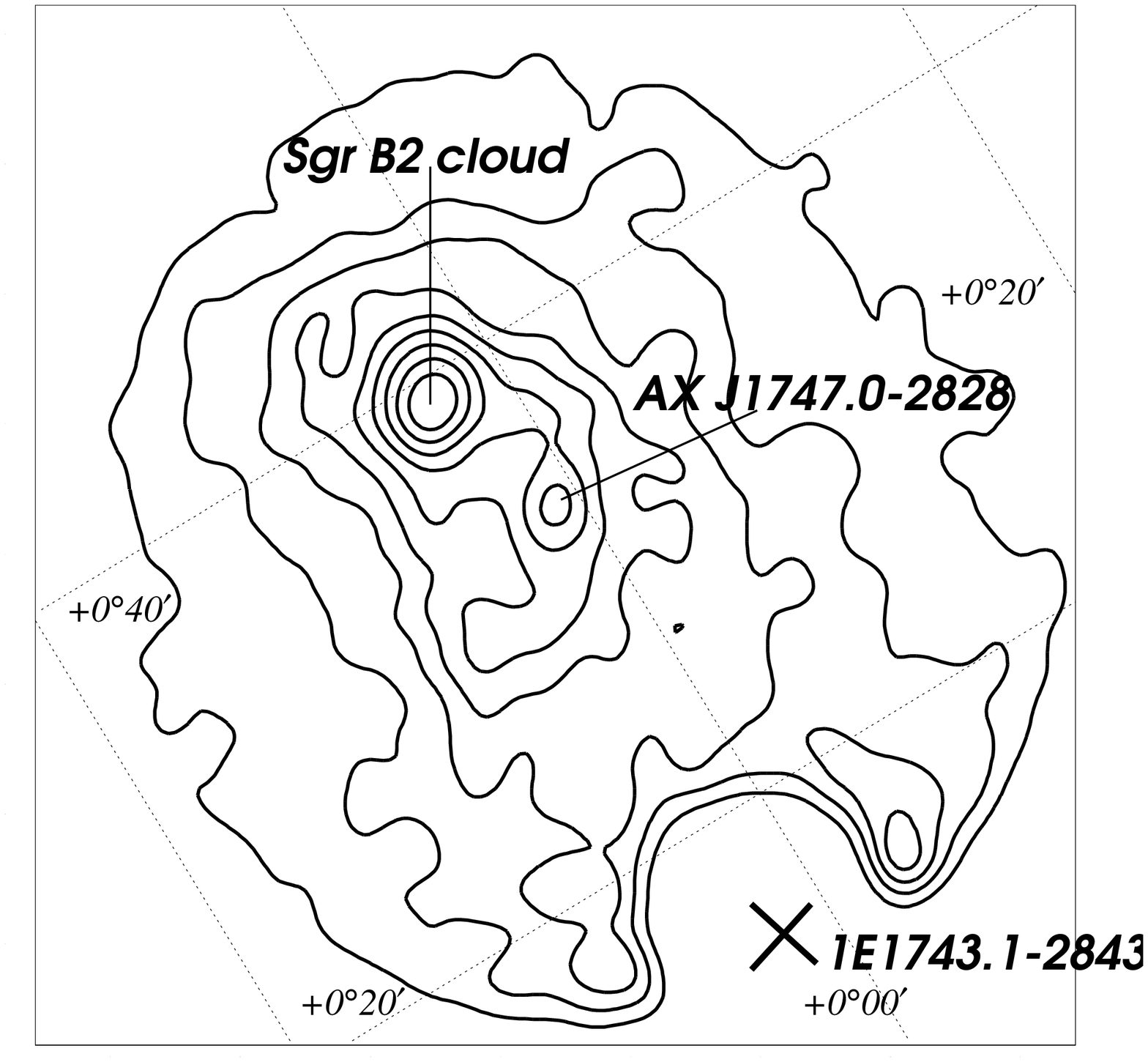,width=8.8cm,clip=} }
\vspace*{-0.5cm}
\centerline{\psfig{file=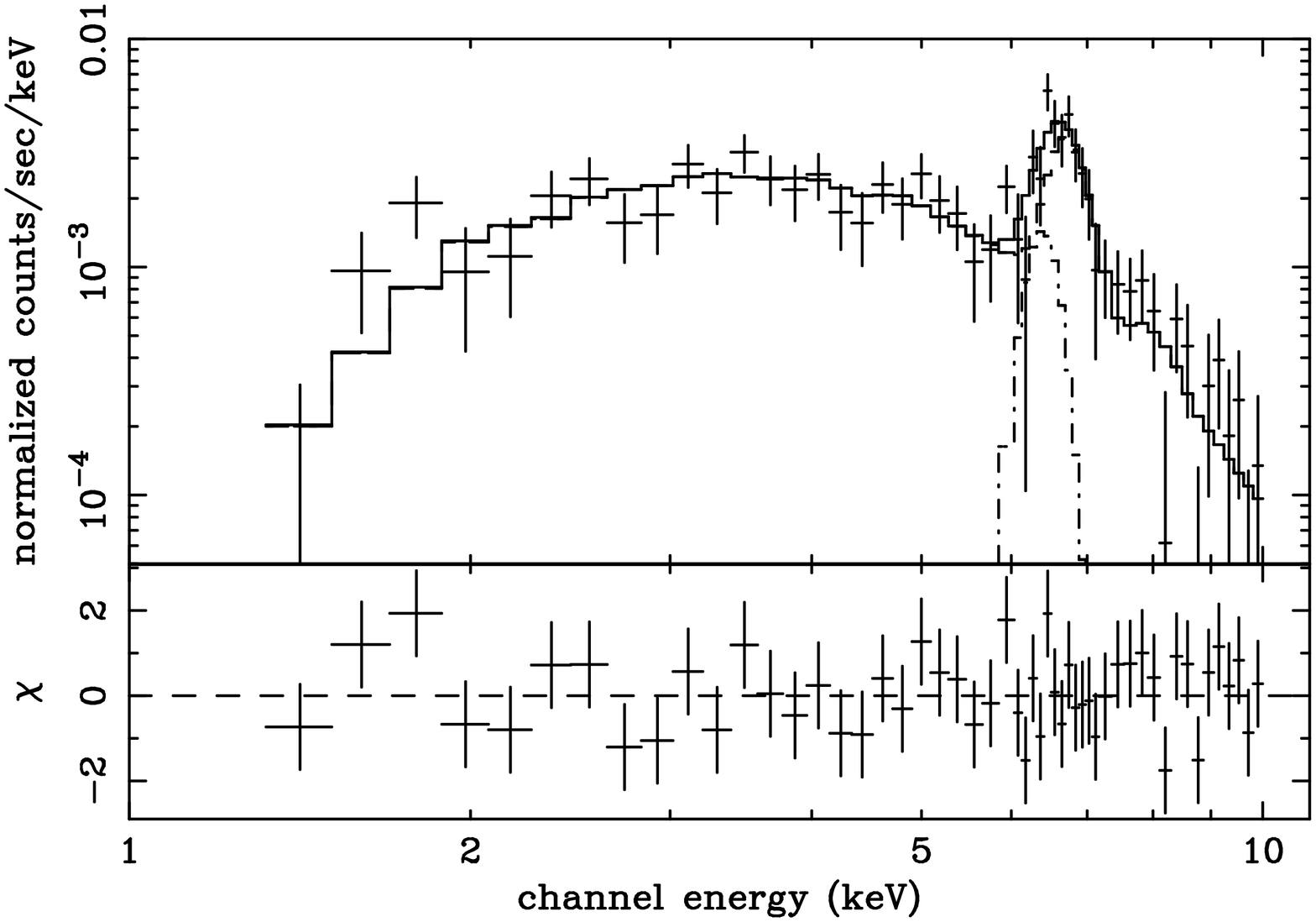,width=8.8cm,clip=} }
\caption{
 (Upper) The same as Fig.~\ref{fig:img:0.9+0.1} upper panel,
 but of G0.56$-$0.01 (AX J1747.0$-$2828) with 6.0--7.0 keV band,
 which is dominated by iron K$\alpha$ line.
  Note that the image was corrected only for exposure, NXB was not subtracted,
 and the region around a bright source 1E~1743.1$-$2843 was excluded
 before the smoothing in order to reduce the contamination from 1E~1743.1$-$2843.
 The positions of the X-ray reflection nebula Sgr~B2 cloud
 (e.g., \cite{Murakami99}) and 1E~1743.1$-$2843
 are also indicated.
(Lower) The same as Fig.~\ref{fig:359.1-0.5} lower panel, but of G0.56$-$0.01.
  The fitting model is the thin thermal plasma model
 and 6.4-keV narrow line, both with interstellar absorption.
\label{fig:G0.56-0.01}}
\end{figure}

   Fig.~\ref{fig:G0.56-0.01} (upper panel) shows the GIS image of G0.56$-$0.01
 (AX~J1747.0$-$2828) in 6--7 keV band, where this source can be seen
 the most significantly.  The X-ray emitting region is compact
 and not resolved with GIS.

   The background subtracted spectrum is given in Fig.~\ref{fig:G0.56-0.01}
 (lower panel).
   We accumulated the source X-ray photons from the $3'$-radius
 circular region around AX~J1747.0$-$2828, and the background photons,
 from an elliptical region with the major
 axis parallel to the Galactic Plane, excluding the $3'$-radius circular
 regions around AX~J1747.0$-$2828 itself and Sgr B2
 (see Murakami {\etal} (1999)).

    The spectrum is found to be characterized with a quite strong line
 at between 6--7 keV, which is also implied from the X-ray image.
  We tried to fit the spectrum with the thermal
 bremsstrahlung and a Gaussian line.  Then we found the equivalent width
 of the line to be quite large, $\sim$2 keV, and the center energy of
 the Gaussian to be 6.63$\pm$0.06 keV, being consistent with K$\alpha$
 line from helium-like iron.  Hence, the spectrum is definitely
 a thin thermal origin with high temperature of several keV or higher.

    We then fitted the spectrum with a thin thermal plasma model.
  The model well represents the total spectral shape.   The best-fit
 temperature is $kT=6.0^{+1.9}_{-1.5}$ keV, the abundance, $Z>2$ solar,
 and the hydrogen column density, \NH$=(6.1^{+1.6}_{-1.1})\,10^{22}$ {\NHUNIT}.
  The flux is 1.6\,10$^{-12}$ {\FLUXUNIT} in 0.7--10 keV band.

    The large column density suggests this source to be located at
 near the Galactic Center.
  Thus, the absorption corrected X-ray luminosity is estimated
 at $\sim$3.6\,10$^{34}$ {\LUMIUNIT} under the assumption of
 the distance of 8.5 kpc.

    This high temperature and the overabundance suggest AX~J1747.0$-$2828 to be
 a possible new candidate of a young SNR.
  The luminosity is also within the range of that of the typical SNR.
  Although the supernova remnant is the most probable source,
 the other possibility, for example, a cataclysmic variable,
 still cannot be excluded (e.g., \cite{Terada99}).
   In any case, it is a new class object in the Galactic Center region,
 which is a good candidate for the origin of the Galactic Center plasma.

\section{Discussion}

    SNRs are one of the possible candidates for the origin
 of the Galactic Center hot plasma.

   With {\ASCA}, we completely surveyed the region of $|l|<1$\degr and $|b|<0.3$\degr,
 where a large portion of SNRs in this region are expected
 to exist, and partially surveyed some areas
 with larger galactic latitude.
    The number of the detected SNRs with {\ASCA}
 is two or three in 7 cataloged SNRs, and two for the new
 candidates in the Galactic Center region.
  On the other hand, the required number to be responsible
 for the Galactic Center plasma is
 about 10$^3$ in the region of $|l|\leq 1$\degr and $|b|\leq 0.5$\degr.
  Therefore, the number of the detected SNRs is quite
 insufficient.

    From recent radio molecular line observations, Hasegawa {\etal} (1998)
 found over 300 shell-like structures, possibly SNRs,
 in the region of $|l|\leq 0.5$\degr, which correspond to
 about 20 shell-like structures along any line of sight in the region.
  Therefore, the number of SNRs may be possibly too much
 to resolve with {\ASCA}.
  Future observations with higher angular resolution would be required
 to solve this problem.

    The spectra of four of the detected five SNRs (or candidates) are
 relatively soft (G359.1$-$0.5, G359.1$+$0.9, and G0.0$-$1.3 (AX~J1751$-$29.6))
 or have no line-like feature (G0.9$+$0.1).  Thus they cannot explain
 the spectrum of the Galactic Center plasma.  On the other hand,
 a new SNR candidate G0.56$-$0.01 (AX~J1747.0$-$2828) is notable
 for the strong iron line similar to that of the Galactic Center plasma.
   Such class of objects may be a key to understand the origin
 of the plasma.  Search for such objects will be strongly encouraged.

\vskip 0.4cm

\begin{acknowledgements}
   The authors express their thanks to all the members of the {\ASCA} team.
   We are grateful to Drs. J. P. Hughes and P. Slane for their valuable
 comments.
   MS and YM acknowledge the support from Research Fellowships of
 the Japan Society for the Promotion of Science.
\end{acknowledgements}

\end{document}